%% file: mo094.tex
\documentclass[iop,showkeys]{emulateapj}

\submitted{Published in the Astrophysical Journal (2011, 742, 21); includes text missing from the third-to-last paragraph}
\begin{document}

\title{The Extreme Overabundance of Molybdenum in Two Metal-Poor Stars}
\author{Ruth C. Peterson$^1$}
\affil{$^1$Astrophysical Advances and UCO/Lick}

\begin{abstract}
We report determinations of the molybdenum abundances in
five mildly to extremely metal-poor turnoff stars using five Mo II lines near 2000\,\AA.
In two of the stars, the abundance of molybdenum is found to be extremely enhanced, 
as high or higher than the neighboring even-$Z$ elements ruthenium and zirconium. 
Of the several nucleosynthesis scenarios envisioned for the 
production of nuclei in this mass range in the oldest stars, a high-entropy wind 
acting in a core-collapse supernova seems uniquely capable of the twin aspects of a high 
molybdenum overproduction confined to a narrow mass range. 
Whatever the details of the nucleosynthesis mechanism, however, this unusual excess suggests 
that very few individual 
nucleosynthesis events were responsible for the synthesis of the light 
trans-Fe heavy elements in these cases, an unexpected result given that both 
are only moderately metal-poor.

\end{abstract}

\section{Introduction}

The relative abundances of elements heavier than iron in low-mass stars of 
low iron abundance bear silent witness to the exploding massive stars in which 
their metal content was created. Because old massive stars have long since evolved, 
their solar-mass counterparts are the only surviving stellar 
relics in which these events have been recorded. 
Their heavy-element abundance distributions, reviewed by \citet{2008ARA&A..46..241S},  
can yield critical diagnostics of the objects and environments that formed the material, 
and the sequences of events that resulted in the buildup of the halo and disk of our Galaxy.

In single stars of metallicity below one-thirtieth solar, [Fe/H] $<$ $-$1.5, the heavy elements 
from barium ($Z$ = 56) onward owe their existence to the r-process (rapid neutron 
capture on seed iron nuclei). Their proportions with respect to one another are 
the same in all metal-poor stars, despite the wide range of two orders 
of magnitude observed in their overall abundance 
with respect to iron. The most probable r-process site is Type II supernovae, 
whose progenitors are short-lived massive stars.
In more metal-rich single stars, [Fe/H] $\geq$ $-$1.5, 
elements begin to appear that are created by the s-process (slow neutron 
capture), in pulsations in intermediate-mass asymptotic giant branch (AGB) stars. 
The AGB evolutionary time of a few 100 Myr suggests a time delay of this order 
in star formation at and above this metallicity. 

More complex is the origin of the lightest trans-Fe elements gallium 
through cadmium ($Z$ = 31 to 48).
These elements have been attributed in varying degrees to the 
p-process \citep[proton capture;][]{RevModPhys.29.547,1976A&A....46..117A,1978ApJS...36..285W}, 
a ``weak'' s-process \citep{1968psen.book.....C,1989RPPh...52..945K}, a ``weak'' 
r-process \citep{1965ApJS...11..121S,1991PhR...208..267C}, a specific ``LEPP'' 
\citep[light element primary process;][]{2004ApJ...601..864T}
such as the $\nu$p process of \citet{2006PhRvL..96n2502F}, and/or the low-entropy 
domain of a neutrino wind above the neutron star formed 
in a Type II supernova \citep[e.g.][]{1999ApJ...516..381F}.

In this paper, we establish and discuss the abundances of the two light trans-Fe elements 
molybdenum and ruthenium (Mo, Ru; $Z$ = 42, 44) in five metal-poor stars whose 
enhancements of heavy r-process elements are mild.
In two of the five stars, HD 94028 and HD 160617, 
the molybdenum abundance is extremely elevated, 
with important ramifications for the 
synthesis process and the number of synthesis events contributing to the light 
trans-Fe elemental abundances.

\section{Current Observational Constraints on Light Trans-Fe Elements}

As reviewed by \citet{2010ppc..conf..379L}, stringent 
constraints of nucleosynthesis models are derived from isotopic 
abundances of meteorites \citep[e.g.][]{2006LPI....37.2041P}. 
These reflect the integral of the products 
of all processes incorporated into the pre-solar nebula -- a single detailed example 
in space and time.

Isotopes of the light trans-Fe elements have proven 
to be among the most difficult to reproduce. Molybdenum is especially problematical; 
its solar-system p-process isotopic fraction of $\sim$25\% is larger than that of any other 
trans-Fe element \citep{2010ppc..conf..379L}.
This is a stumbling block for models invoking the s-process in low-mass AGB stars 
\citep[e.g.][]{2003ApJ...593..486L}. Such a scenario 
is unlikely to apply in any case to the low-metallicity stars of the halo, 
as it relies on the pre-existence of 
AGB stars of solar metallicity and also of quite low mass, 
1.5\,M$_{\sun}$, with accordingly long main-sequence lifetimes. 
In contrast, 
\citet{1996ApJ...460..478H} succeeded in directly producing light p-process nuclei 
with specific choices of entropy $S$ and electron fraction $Y_e$ in a neutrino-driven wind.
They noted that this is a primary process, one in which 
``the r-process and some light p-process nuclei may be coproduced.''

Recently, \citet{2009PASA...26..194F} reproduced all seven of the solar isotopes of 
molybdenum by selecting models from a parameterized grid of calculations based on 
a high-entropy wind (HEW) operating in Type II supernovae. They 
find it ``can co-produce the light p-, s-, and r-process isotopes 
between Zn ($Z$ = 30) and Ru ($Z$ = 44) at electron abundances in the range 0.450 $\leq$ 
$Y_e$ $\leq$ 0.498 and low entropies of $S$ $\leq$ 100 -- 150. Under these 
conditions, the light trans-Fe elements are produced in a charged-particle 
($\alpha$-) process, including all p-nuclei up to $^{96,98}$Ru. In our model, 
no initial SS [solar system], s- or r-process seed composition is invoked; 
hence, this nucleosynthesis component is primary.'' 
In part because ``the overall yields of the light trans-Fe elements decrease with 
increasing $Y_e$'', they conclude that ``more quantitative
answers to questions concerning the astrophysical site of the compositions of the
LEPP elements between Sr ($Z$ = 38) and Cd ($Z$ = 48), as well as all of the n-capture
elements, will require more and higher quality observational data and also more
realistic values of entropy superpositions derived from hydrodynamical models.''

Abundances of lighter and heavier elements 
in metal-poor stars are already providing further constraints.
\citet{2010ApJ...724..975R} derived abundances for zinc, yttrium, lanthanum, europium,
and lead (Zn, Y, La, Eu, and Pb; $Z$ = 30, 39, 57, 63, and 82) in 161 metal-poor stars 
with [Fe/H] $<$ $-$1.4. Based on models of the s-process in AGB stars, they used [Pb/Fe] 
to identify stars with no discernible s-process contribution, and concluded that 
s-process elements were largely absent from progenitor material at these low metallicities.
Because a scatter remained in [La/Fe] in those stars 
with relatively low r-process content, they confirmed the result emphasized earlier by 
\citet{2007ApJ...666.1189H}, that the ratio of light r-process to heavy r-process 
elements varies widely among metal-poor stars. 
\citet{2010ApJ...724..975R} also confirmed an anti-correlation between Y/Eu and Eu/Fe 
\citep{2007A&A...476..935F}, and showed that Y production was decoupled 
from both Zn and Fe. They were able to reproduce the range of Y/Eu ratios with 
simulations of HEW models that explore the effects of a range of entropies 
\citep{2010ApJ...712.1359F}.

For molybdenum itself, previous abundance determinations in metal-poor stars 
are restricted to giants and subgiants, in which near-UV and optical  Mo I lines are 
detectable.  Except for giants with extreme r-process enhancements 
\citep[e.g.][]{2003ApJ...591..936S}, published [Mo/Fe] values are all near solar. 
Table~\ref{tbl-mo} summarizes values for [Mo/Fe], and other light trans-Fe elements 
where available, for eight field halo stars with 
[Fe/H] $\leq$ $-$1.4 and [Eu/Fe] $<$ 0.9, as an indicator of mild r-process enhancement.
Of the 16 globular-cluster studies listed in Table 1 of \citet{2011ApJ...732L..17R}, 
only one presents results for Mo. In that study, for eight giants in the globular 
cluster M5, \citet{2011AJ....141...62L} find [Fe/H] = $-$1.43, [Eu/Fe] = +0.49,
[Zr/Fe] = +0.34, and [Mo/Fe] = $-$0.10, with no significant star-to-star variation. 
The referee adds that \citet{2008ApJ...689.1031Y} have derived [Mo/Fe] for 11 stars 
in M4 and two in M5, none of which has [Mo/Fe] $>$ +0.4. Since M4 has a 
subgroup of stars with significant s-process contribution 
\citep[Fig.~1, panel 4 of][]{2011ApJ...732L..17R}, 
its non-s-process [Mo/Fe] upper bound may even be lower.
Among the two dozen normal field and cluster giants in which molybdenum has been 
studied to date, then, none has [Mo/Fe] $>$ +0.4.

\input{table_mo.tex}

\section{Stellar Spectra} 
In this work we provide additional support and constraints for HEW production of light 
trans-Fe elements, by determining the abundances of Mo and Ru in metal-poor turnoff stars 
from Mo II and Ru II lines near 2000\AA\ in high-resolution spectra taken with the Space 
Telescope Imaging Spectrograph (STIS). Five such spectra were found in the MAST archive, 
the Multimission Archive at the Space Telescope Science Institute (STScI). 
To constrain the molybdenum abundance scale and to derive abundances for other elements, 
archival near-UV and optical echelle spectra were analyzed for the same stars. 

Table~\ref{tbl-sp} lists for each spectral 
region the spectra employed for each star. 
Reductions by others were adopted from StarCat \citep{2010ApJS..187..149A}, 
the UVES pipeline\footnote{http://www.eso.org/sci/software/pipelines/},
Keck HIRES archival extractions, 
and the UVES ground-based spectral programs of the Next Generation Spectral 
Library \citep{2004AAS...205.9406G}. Our own reductions used the IRAF\footnote
{IRAF is distributed by the National Optical Astronomy Observatories, which are operated 
by the Association of Universities for Research in Astronomy, Inc., under cooperative 
agreement with the National Science Foundation.} environment. 
We performed bias and dark removal, 
coadded multiple spectral images of the same object with cosmic-ray removal, 
extracted orders with removal of sky and local interorder background, 
corrected the dispersion using Th-Ar exposures, 
and rectified the continuum and spliced together adjacent orders. 

\input{table_spectra.tex}

\section{Synthetic Spectral Analysis}
We have derived stellar parameters and abundances by matching each stellar spectral observation to 
theoretical spectra calculated for each star using an updated version of 
the SYNTHE program of \citet{1993KurCD..18.....K}. We input a list of molecular and atomic 
line transitions with wavelengths, energy levels, and gf-values, and a model atmosphere 
characterized by effective temperature $T_{{\rm eff}}$, surface gravity log {\it g}, 
microturbulent velocity $V_{{\rm t}}$, and logarithmic iron-to-hydrogen ratio [Fe/H] 
with respect to that of the Sun. Our models are interpolated in the grid of 
\citet{2003IAUS..210P.A20C}.

The line lists are based on 
the Kurucz \emph{gfhy}\footnote{http://kurucz.harvard.edu/LINELISTS/GFHYPER100/}
atomic lines with known energy levels (``laboratory'' lines), along with 
Kurucz diatomic molecular line lists\footnote
{http://kurucz.harvard.edu/LINELISTS/LINESMOL/} and TiO lines\footnote
{http://kurucz.harvard.edu/molecules/TiO/} \citep{1998FaDi..109..321S}.
We have modified these extensively in the near-UV and optical, 2200\AA\ -- 9000\AA, 
by matching calculations to echelle spectra of standard
stars.  Starting with weak-lined stars, we calculated each spectrum, adjusted 
gf-values singly for atomic lines and as a function of band and energy for molecular 
lines, and guessed identifications of ``missing'' lines, those present in the spectra 
but not in the laboratory line list, which 
become an extreme problem in the UV. 
\citet{2001ApJ...559..372P} detail the procedure, 
and that work and \citet{2008STScINews...4.1P} show 
that our calculations agree well with observed optical and
mid-UV spectra of nearby mildly metal-poor stars. 

\input{table_stellar.tex}

Following these procedures, 
we continued the line modifications into the 2000\AA\ region.
This better defines the local continuum, especially 
in the two stronger-lined stars HD 76932 and HD 211998. 
All missing lines were simply assumed to be
${\rm Fe\,{\textsc{i}}}$ lines with lower excitation
potential below 1 eV, as these are extremely common among the identified lines.
We added these at wavelengths where the spectra of
HD 76932 and HD 211998 showed absorption at the same wavelength
that was not matched by any laboratory line in the Kurucz atomic line lists,
even when its gf value is increased by as much as 2.5 dex.

We find stellar parameters strictly from the spectra, and not from colors.
Effective temperature $T_{{\rm eff}}$ comes from demanding that the same abundance 
emerge from low- and high-excitation lines of same species (usually 
${\rm Fe\,{\textsc{i}}}$). 
Gravity log {\it g} comes from the wings of other strong lines. 
Demanding no trend in abundance with line strength sets microturbulent velocity $V_{{\rm t}}$.
Iron abundance [Fe/H] follows by matching relatively unblended weak lines, 
as do the abundances of other elements. 
The resulting uncertainties are typically 0.1 -- 0.2\,dex in [X/Fe] for element X, 
if represented by at least three lines whose blending, if any, is reliably modeled, 
and whose gf-values are well-determined.
Comparing ${\rm Fe\,{\textsc{i}}}$ and ${\rm Fe\,{\textsc{ii}}}$ abundances 
confirms or refines gravity, and the wings of the Balmer lines confirm $T_{{\rm eff}}$. 
These agree with other $T_{{\rm eff}}$ diagnostics only when convective overshoot is turned 
off, as is true of the \citet{2003IAUS..210P.A20C} models but not those of 
\citet{1993KurCD..13.....K}.
Table~\ref{tbl-st} lists the resulting stellar model parameters and abundances.

Figures~\ref{fig:fig-moru} and \ref{fig:fig-moruhi} compare the calculations 
based on these parameters 
to the observations around selected Mo and Ru lines.
Figure~\ref{fig:fig-moru} extends the region surrounding each line, to 
indicate the overall goodness of fit and the choice of continuum placement. 
Figure~\ref{fig:fig-moruhi} expands the scale in the vicinity of each line, to
portray the effect of a $\pm$0.3 change in abundance of the element in question.

\section{Molybdenum and Ruthenium Abundances}

\begin{figure*}[ht!]
\epsscale{1.16}
\plotone{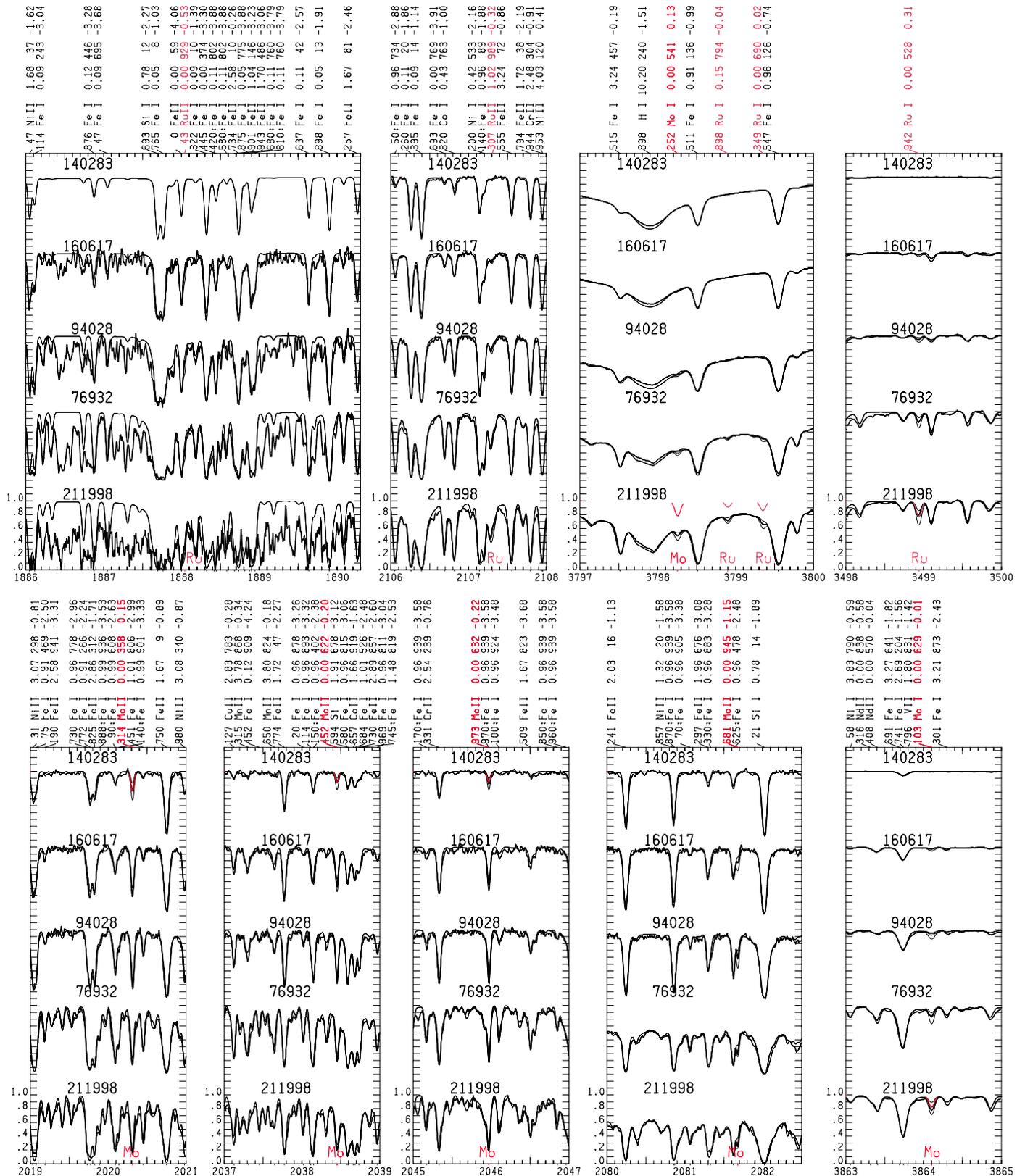}
\caption{
Comparisons are shown between observed and calculated spectra
in nine spectral regions, indicated by 
wavelength in \AA ngstroms at the bottom. 
Plots for the five individual stars are offset vertically; 
ticks on the y axis indicate one-tenth of the normalized continuum level.
The HD number of each star is given above its plot.
The heavy line is its observed spectrum, and the light line
its calculated spectrum. The strongest lines in the calculated 
spectrum are identified at the top. First are the digits following 
the decimal place of the line center wavelength (in vacuum for 
the bluest region, and in air otherwise). Next is given 
the species giving rise to the line; a colon indicates 
a ``missing'' line whose identification was assumed to be Fe I. 
Following this are the lower excitation potential of the line in eV, 
an indicator of its strength, and its log gf-value. 
Three calculations are shown near Mo and Ru lines; these are
expanded and described in Figure~\ref{fig:fig-moruhi}.
}
\label{fig:fig-moru}
\end{figure*}

\begin{figure*}[ht!]
\epsscale{1.04}
\plotone{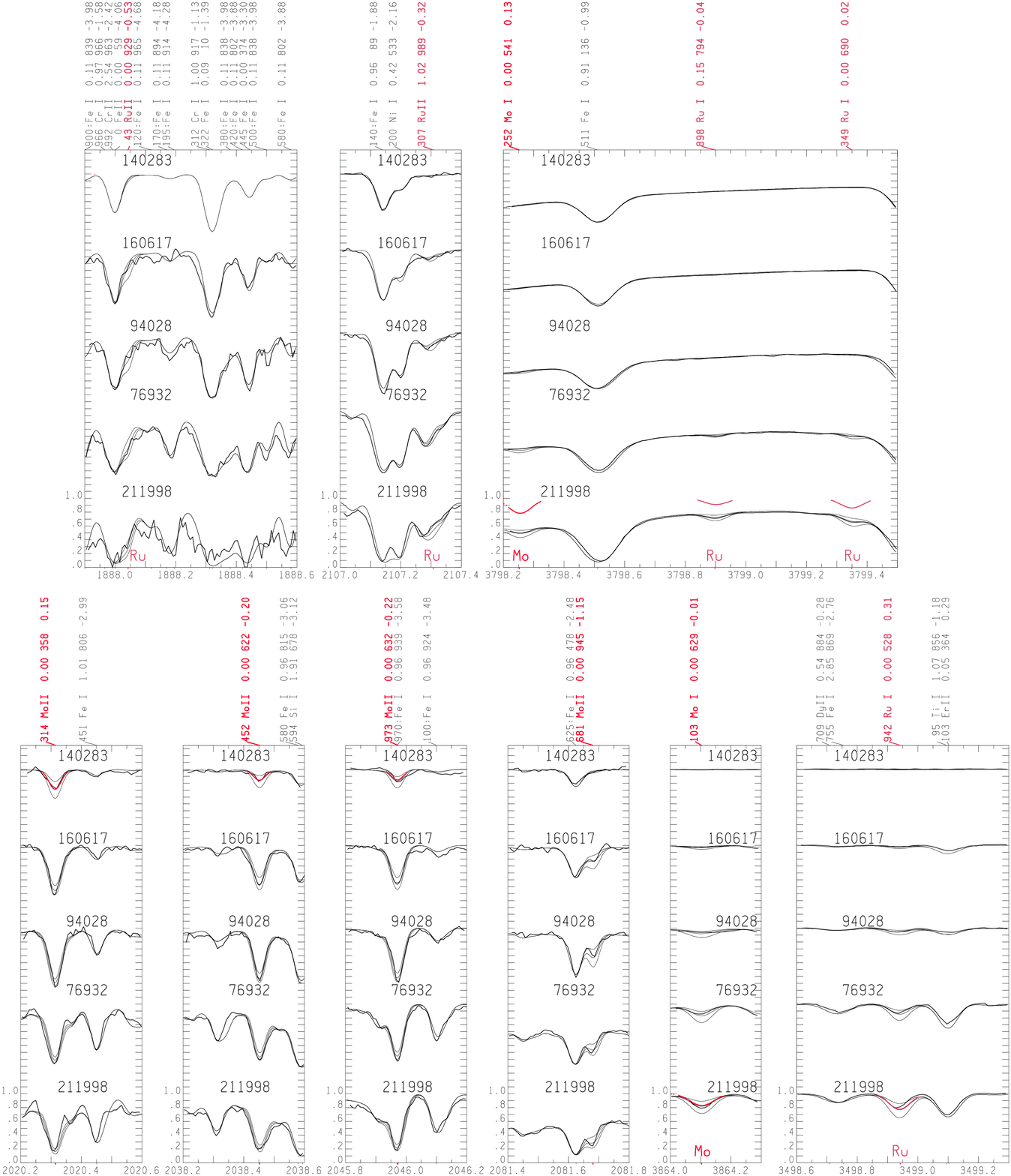} 
\caption{
The observed and calculated spectra of Figure~\ref{fig:fig-moru} 
are shown in the same nine spectral regions,  
and with the same calculations and labels, but 
on an expanded wavelength scale in the vicinity of the Mo and Ru lines. 
The middle light line is calculated assuming 
the abundances of Table~\ref{tbl-st}. The exception is the choice 
of [Ru/Fe] = +0.1 $\pm$0.3 for HD 140283, 
as no Ru I nor Ru II lines with reliable gf-values were detected 
for this star. The weaker and stronger light 
lines in each plot indicate calculations with Mo and Ru abundances 
0.3\,dex lower and higher.  Red arcs highlight a few cases 
in which the line indicated was the only line calculated.
}
\label{fig:fig-moruhi}
\end{figure*}

As seen in these figures, the calculated spectra match the fitted 
spectra quite well. However, the abundance determinations are 
affected by systematic errors, due to the uncertainties in continuum 
placement and in gf-values. These are difficult to judge.

Continuum placement is affected by line absorption, 
whose modeling depends on gf-value adjustment and on correct 
assignment of the wavelength, species, and lower excitation potential 
of missing lines. The number of missing lines grows dramatically 
toward the UV, and consequently the continuum becomes less well defined 
below 2000\AA, even for the weak-lined star HD 160617. This is 
illustrated in the top left panel in Fig.~\ref{fig:fig-moru}.
Line-strength adjustment was attempted only 
over 1887.6\AA\ -- 1889.0\AA, as  
spectral data are currently lacking for the weakest-lined star 
HD 140283, and many significant lines are missing. 
The latter shortcoming is likely to be alleviated by revising
the input line list to include recent recalculations 
by \citet{2011CaJPh..89..417K}, posted on the Kurucz web site in April 2011, 
which include ten times as many lines of Mg, Si, Ca, and the iron-peak elements. 

Their inclusion might change the Ru abundances of HD 160617 and HD 94028 by 
up to 0.3 dex. While the blending of the Ru II line itself can be modeled 
primarily with identified lines, if missing lines are also 
present but currently unrecognized, 
the currently inferred ruthenium abundance may be overestimated. However, 
if there are more missing lines in the adjacent regions used 
to set the continuum, its level is underestimated, and the 
inferred ruthenium abundance may be underestimated.

Scale errors can occur in both laboratory 
and theoretical gf-values. We examined these wherever possible. 
To calculate the UV Mo II lines, we increased by 0.133 dex the Mo II log gf values of
\citet{2001JPhB...34..477S} for the five Mo II transitions that originate from the
ground state. This is the difference between their value and the recent result of
\citet{2010JPhB...43h5004L} for the 2082\AA\ Mo II line, the only ground-state
Mo II line in common between the two studies. We followed
\citet{2006ApJ...645..613I} in adopting the Mo I gf-values of \citet{1988PhyS...38..707W}
and the Ru I results of \citet{1994JQSRT..51..545W}. For Ru II, \citet{1994ApJ...421..809J}
provide 18 experimental gf-values, and \citet{2009JPhB...42p5005P} give theoretical values. Both wavelengths and gf-values 
proved unreliable for ruthenium lines with data solely from other sources.

For HD 76932 and HD 211998, the molybdenum abundances were determined from
the optical Mo I line at 3864\AA.
For HD 94028, we adopted the molybdenum abundance which was just consistent
with the non-detection of the Mo I 3797\AA\ line.
For HD 160617 and HD 140283, the molybdenum abundances were determined by matching
Mo II line strengths, as these lines are rather weak, and Mo I is not detected.
The Mo I and Mo II gf-value scales thus appear to be on a consistent scale.
This is encouraging, as the Mo I gf-values have yielded a solar molybdenum
abundance \citep{1983ApJ...275..889B} within 10\% of the meteoritic
value \citep{2010ppc..conf..379L}. 

For ruthenium, we sought the strongest Ru I lines in the optical and Ru II 
lines in the UV. It is again encouraging 
that Ru I and Ru II lines give consistent results for HD 76932 and HD 211998.
For the other stars, however, the ruthenium lines with gf-values from 
these two sources that lie within the wavelength ranges observed at high 
resolution are weak and blended. 
In HD 160617 and HD 94028, our Ru values are based on these plots of 
the two Ru II lines and the Ru I line at 3499.942\AA. 
For HD 140283, the limit is set by calculations not shown, 
adopting [Ru/Fe] = +1.0 for the Ru II 2102.307\AA\ line. 

\section{Cadmium}

Cadmium ($Z$ = 48) is currently constrained in only two stars. 
In HD 94028, [Cd/Fe] $\sim$ 0.0 is found from lower-resolution E230M spectra 
of the Cd I 2288.018\AA\ line, 
adopting log gf = +0.15\,dex \citep{2010ApJ...714L.123R}. 
The HD 140283 E230H spectra show artifacts in this region; 
data are lacking entirely for other stars. In HD 140283, 
[Cd/Fe] $<$ 0.0 is inferred from the Cd II 2144.393\AA\ line, 
even with log gf = $-$0.11 
\citep{1973JQSRT..13..369A}, which is lower than more recent gf-values 
for this line \citep[$-$0.04 -- +0.12:][]{2004PhRvA..70d2508X, 2005PhyS...72..142M}. 
In the other stars, this line is possibly blended. 
Obtaining E230H data for the Cd I line at 2288.018\AA\ for these stars 
should pinpoint cadmium abundances, 
by resolving its own potential blends, determining [Cd/Fe] in HD 140283 and 
so fixing the Cd II 2144.393\AA\ gf-value, and then using Cd I and Cd II together 
to set [Cd/Fe]. This would more strongly constrain the extent of light trans-Fe 
elemental overabundances in atomic number $Z$, and so the models of their 
production.

\section{The s-process Contribution to Stellar Heavy-Element Abundances}

The s-process may contribute to the Ru, and Mo, abundances, especially 
for the more metal-rich stars.
To check this we have included [La/Fe] abundances from the 3995.75\AA\ and 4086.71\AA\
La II lines, with gf-values from \citet{2001ApJ...556..452L} 
and hyperfine splitting structure from \citet{2006ApJ...645..613I}. 
We compare our Table~\ref{tbl-st} values with the trend of [La/Fe] vs.\ [Eu/Fe] 
for r-only field halo stars in the fourth panel of Fig.~1 of \citet{2011ApJ...732L..17R}. 
HD 76932 and HD 211998 lie within 0.05\,dex of the mean relation. 
HD 160617 and HD 94028 fall $<$ 0.3 dex above the line, marginally beyond 
the extent of the r-only field stars in that figure. We conclude that any excess 
s-process contribution is small in these four stars. 

For HD 140283, none of these La II lines were detected, but at [Fe/H] = $-$2.6  
an s-process contribution is unlikely \citep{2010ApJ...724..975R}. 
\citet{2010A&A...523A..24G} find ambiguous results for its s-process content 
from fitting the 4554\AA\ and 4934\AA\ Ba II line profiles. That work and ours 
agree that [Eu/Fe] $<$ $-$0.9 in this star. Its abundance pattern strongly resembles 
that of HD 88609 and HD 122563 in showing larger deficiencies of heavy 
than light trans-Fe elements, with low relative abundances overall 
\citep{2007ApJ...666.1189H}. 

\section{Results and Implications for Nucleosynthesis}

Thus we conclude that in HD 94028, molybdenum is extremely enhanced, 
more so than ruthenium. Mo is also highly enhanced in HD 160617. 
Both these stars, indeed four of the five metal-poor turnoff stars studied here, 
have higher [Mo/Fe] values than any of the eight metal-poor giants with mild 
(if any) r-process enhancements listed in Table~\ref{tbl-mo}. Yet none has 
[La/Fe] enhancements more than 0.3\,dex higher than expected from its [Eu/Fe] 
value, based on Fig.~1 of \citet{2011ApJ...732L..17R}. The total range in 
[Mo/Fe] among these five stars is 0.8\,dex, while the total range in [La/Fe] 
is 0.4\,dex. Among these stars with modest or no r-process enhancements, 
no correlation is seen between 
[Eu/Fe] and [Mo/Fe]. The extreme enhancements 
are rather narrowly confined to Mo, diminishing towards Zr and Y and 
towards Ru and beyond.

Since both s- and r-process nucleosynthesis tend to produce similar enhancements over a 
range of non-magic neighboring even-$Z$ elements, the production of molybdenum and ruthenium 
in HD 94028 most probably involves another process. As noted above, a 
HEW is capable of overproducing light trans-Fe elements if the 
wind parameters are right. \citet{2010ApJ...712.1359F} express this in terms of $Y_e$: 
$Y_e$ = 0.498 yields Sr, Y, Zr, and Nb ($Z$ = 38--41); $Y_e$ = 0.496, Mo and Ru; 
$Y_e$ = 0.490, Rh, Pd, and Ag (Z = 45--47); and $Y_e$ = 0.482, Cd and beyond (Z $\geq$ 48). 
Clearly, a HEW with a limited parameter range seems 
able to reproduce the strong excess of molybdenum with less strong 
excesses at zirconium and beyond ruthenium.

These unique factors also suggest that very few individual nucleosynthesis events 
were incorporated into the stars with extreme molybdenum abundances. This is especially 
remarkable for HD 94028, given its rather high metallicity, [Fe/H] = $-$1.4. 
From Table~7 of \citet{2010ApJ...712.1359F}, the yield from an individual HEW event 
with $Y_e$ = 0.496 is $10^{-5}$\,M$_{\sun}$; this is more than adequate. However, the narrow 
entropy range means that multiple HEW events with a range of parameters must be 
avoided during the buildup of the iron abundance. 

The referee has pointed to another example, from 
\citet{2007PASJ...59L..15A}. They find that the star COS 82 in the dwarf galaxy 
Ursa Minor, with [Fe/H] = $-$1.5, has a very high heavy element enhancement 
with an r-process signature. They note that such high r-process enhancements 
are found in Galactic stars only below [Fe/H] = $-$2.5, 
and suggest that ``the neutron-capture elements
of COS 82 might be provided by a single event.''

Such a scenario has recently become theoretically more feasible. 
Bland-Hawthorn, Sutherland, \& Karlsson (ApJ, in preparation) 
have found that a dark matter halo of about $3 \times 10^6 M_{\sun}$ 
with about $10^5 M_{\sun}$ in gas, 
a lower-mass system than previously thought, 
can still form stars from retained 
products released by an exploding supernova. The lower mass 
results from the reduced sweeping of products when clumpiness of the 
medium and off-center supernovae explosions are included.

However, as discussed above, we know of no prior 
evidence for single-supernova production of any group of trans-Fe elements 
in moderately metal-poor Galactic halo stars. More 
Mo abundances in halo stars with [Fe/H] $\leq$ $-$1.4 are needed, to verify 
rarity and define the frequency of occurrence of high [Mo/Fe] values as a function 
of metallicity. A survey that includes stars with well-established kinematics 
might reveal if high-Mo stars might have been formed in captured/dissipated 
dwarf galaxies. For stronger-lined turnoff stars like HD 76932 and HD 211998,
either cool or of moderate metallicity, Mo and Ru can be determined from archival
ground-based spectra, as shown in Figure~\ref{fig:fig-moru}. For the most 
metal-poor r-normal stars, however, the UV lines are required. 

Determination of the abundances of more light trans-Fe elements in 
HD 94028 and HD 160617 seem critical as well, to provide stronger
HEW support and constraints for parameterized modeling. The steeper the
falloff of abundance enhancements from Mo to Ru and beyond, the more narrow
the entropy range implied in HEW synthesis. Obtaining 
high-resolution spectra redward of 2200\AA\ for these stars might reveal 
Nb and Pd abundances, as well as pin down Ru and Cd. 

Both these efforts are vital in revealing whether the very high molybdenum 
enhancements reported here are unique. This in turn would help establish to what
extent they truly are the result of very few events in some stars, even at quite 
high metallicity.

\acknowledgements 
We thank M. Spite for insightful discussions, J. Bland-Hawthorn for describing 
his recent work, J. X. Prochaska for providing his reductions of the Keck HIRES 
data, 
and  D. Silva and R. Hanuschik for providing the reduced UVES NGSL spectra. 
We also appreciate the helpful report of the referee.
Ground-based spectra are based on
observations made with ESO Telescopes at the Paranal Observatory 
with the UVES spectrograph under programs 065.L-0507(A), 072.B-0585(A), and 266.D-5655(A),
and with the Keck Observatory HIRES spectrograph, under programs H6aH (PI A. Boesgaard) 
and U35H (PI A. Wolfe).
This research has made use of the Keck Observatory Archive (KOA), which is 
operated by the W. M. Keck Observatory and the NASA Exoplanet Science Institute 
(NExScI), under contract with the National Aeronautics and Space Administration.
Space-based spectra are based on observations made with the NASA/ESA Hubble 
Space Telescope under GO programs 7348, 7402, 8197, 9455, 9491, and 9804. 
These data were obtained from the HST and StarCat archives hosted 
by the Multimission Archive at the Space Telescope Science Institute (MAST). 
STScI is operated by the association of Universities for Research in Astronomy, 
Inc.\ under NASA contract NAS 5-26555. 

\bibliography{mo094}

\end{document}

%% file: table_mo.tex
\begin{deluxetable*}{rccccccccccccc}
\tabletypesize{\scriptsize}
\tablecaption{
Abundance Determinations for Light Trans-Ironic Elements 
in Mildly r-Process-Enhanced Field Halo Stars\label{tbl-mo}}
\tablehead{
\colhead{Star} 
& \colhead{$T_{{\rm eff}}$}
& \colhead{log {\it g}}
& \colhead{$V_{{\rm t}}$}
& \colhead{[Fe/H]}
& \colhead{[Eu/Fe]}
& \colhead{[Zr/Fe]}
& \colhead{[Nb/Fe]}
& \colhead{[Mo/Fe]}
& \colhead{[Ru/Fe]}
& \colhead{[Rh/Fe]}
& \colhead{[Pd/Fe]}
& \colhead{[Ag/Fe]}
& \colhead{[Cd/Fe]}
} 

\startdata
HD  88609  & 4550 & 1.10 & 2.40 & $-$3.07 
& $-$0.33 & +0.24 & $-$0.07 & +0.15 & +0.32 & $<$+0.70 & +0.03 & +0.10 & ... \\

HD 122563  & 4570 & 1.10 & 2.20 & $-$2.77 
& $-$0.52 & $-$0.10 & $-$0.13 & $-$0.02 & +0.07 & $<$+0.45 & $-$0.28 & $-$0.05 & $-$0.5 \\

HD 221170  & 4510 & 1.00 & 1.80 & $-$2.18
& +0.80 & +0.25 & +0.39 & +0.29 & +0.56 &    +0.71 & +0.46 & +0.44 & ... \\

BD+10 2495 & 4710 & 1.30 & 1.55 & $-$2.45 
& +0.13 & +0.00 & $<$+1.46 & +0.02 &  ...  &     ...  &  ...  &  ...  & ... \\

BD+29 2356 & 4760 & 1.60 & 1.45 & $-$1.55 
& +0.41 & +0.34 & $<$+1.78 & +0.19 &  ...  &     ...  &  ...  &  ...  & ... \\

BD+30 2611 & 4330 & 0.60 & 1.85 & $-$1.45 
& +0.65 & +0.01 &  ...  & +0.02 &  ...  &     ...  &  ...  &  ...  & ... \\

HD 128279  & 5050 & 2.35 & 1.50 & $-$2.45 
& $-$0.25 & $-$0.20 & $<$+1.24 & +0.20 &  ...  &     ...  &  ...  &  ...  & ... \\

HD 175305  & 4770 & 1.80 & 1.25 & $-$1.60 
& +0.35 & +0.23 & $<$+1.32 & +0.06 &  ...  &     ...  &  ...  &  ... & ...  

\enddata
\tablecomments{Units: $T_{{\rm eff}}$, $\mathrm{K}$; $V_{{\rm t}}$, {km~s\ensuremath{^{-1}}}. Atomic numbers: Eu, 63; Zr, 40; Nb, 41; Mo, 42; Ru, 44; Rh, 45; Pd, 46; Ag, 47; Cd, 48.
References: HD 88609,  \citet{2007ApJ...666.1189H}; HD 122563, \citet{2006ApJ...643.1180H}, 
except Cd, \citet{2010ApJ...714L.123R}; 
HD 221770, \citet{2006ApJ...645..613I}; remaining stars, \citet{2010ApJ...711..573R}.
}
\end{deluxetable*}

%% file: table_spectra.tex
\begin{deluxetable*}{rcccccc}
\tabletypesize{\scriptsize}
\tablecaption{Spectral Observations\label{tbl-sp}}
\tablehead{
\colhead{Star} 
& \colhead{Wavelength(\AA)} & \colhead{Instrument} & \colhead{Program} & \colhead{Date (UT)} & \colhead{Time (ks)} & \colhead{Reduction}
}

\startdata
HD 140283 & 1950 -- 2300 & STIS E230H & GO 7348 & 1999-04-09 & 18.32 & StarCat uvsum2126  \\
 & 2378 -- 2891 & STIS E230H & GO 9455 & 2002-08-22 & ~5.28 & IRAF  \\ 
 & 2885 -- 3147 & STIS E230H & GO 9491 & 2003-07-11,12,13,16,17,22,23,24 & 62.57 & StarCat 52831-52844  \\ 
 & 3080 -- 5953 & HIRES & U35H & 2005-03-17 & ~0.60 & HiRedux  \\

HD 160617 & 1880 -- 2150 & STIS 230H &  GO 8197 & 1999-10-29,30; 2000-03-15,16;08-31 & 39.39 & StarCat 51480-51787  \\
 & 3057 -- 3873 & UVES &  65.L-0507(A) & 2000-04-09 & ~3.00 & Pipeline  \\
 & 4400 -- 6780 & HIRES & H6aH         & 2000-05-28 &  0.42 & Extracted \\

HD 94028  & 1880 -- 2150 & STIS 230H & GO 8197 & 2000-05-16,21,26 & 33.05 & IRAF  \\
 & 2278 -- 3120 & STIS 230M & GO 7402 & 1998-12-18       & ~0.60 & IRAF  \\
 & 3050 -- 4989 & UVES & 072.B-0585(A) & 2004-03-10 & ~0.75 & NGSL  \\

HD 76932  & 1880 -- 2150 & STIS 230H &  GO 9804 & 2004-02-19,21 & 23.86 & StarCat 53054-53056  \\
 & 3022 -- 4975 & UVES & 266.D-5655(A) & 2001-03-14 & ~0.34 & Pipeline  \\

HD 211998 & 1880 -- 2150 & STIS 230H & GO 9804 & 2004-08-26,27 & 29.40 & IRAF  \\
 & 3040 -- 10400 & UVES & 266.D-5655(A) & 2002-02-09 & ~0.60 & Pipeline  

\enddata
\end{deluxetable*}

%

%% file: table_stellar.tex
\begin{deluxetable*}{rcccccccccc}
\tabletypesize{\scriptsize}
\tablecaption{Stellar Parameters and Light-Element Abundances\label{tbl-st}}
\tablehead{
\colhead{Star} 
& \colhead{$T_{{\rm eff}}$}
& \colhead{log {\it g}}
& \colhead{[Fe/H]}
& \colhead{$V_{{\rm t}}$}
& \colhead{[Eu/Fe]}
& \colhead{[Y/Fe]}
& \colhead{[Zr/Fe]}
& \colhead{[Mo/Fe]}
& \colhead{[Ru/Fe]}
& \colhead{[La/Fe]}
} 
\startdata
HD 140283 & 5700 & 3.6 & $-$2.6 & 1.3 & $<$ $-$0.9 & $-$0.4 & $-$0.1
& +0.2 & $<$ +1.0 & ... \\

HD 160617 & 6000 & 3.8 & $-$1.8 & 1.2 & +0.6 & +0.0 & +0.4 
& +0.8 & +0.6 & +0.24 \\

HD 94028  & 6050 & 4.3 & $-$1.4 & 1.2 & +0.3 & +0.2 & +0.5 
& +1.0 & +0.7 & +0.30 \\

HD 76932  & 5900 & 4.1 & $-$0.9 & 1.2 & +0.4 & +0.0 & +0.2 
& +0.6 & +0.4 & +0.14 \\
 
HD 211998 & 5300 & 3.1 & $-$1.5 & 1.5 & +0.2 & +0.2 & +0.5 
& +0.5 & +0.5 & $-$0.06  

\enddata
\tablecomments{Units: $T_{{\rm eff}}$, $\mathrm{K}$; $V_{{\rm t}}$, {km~s\ensuremath{^{-1}}} 
}
\end{deluxetable*}

%% file: mo094.bbl
\begin{thebibliography}{50}
\expandafter\ifx\csname natexlab\endcsname\relax\def\natexlab#1{#1}\fi

\bibitem[{{Andersen} \& {Soerensen}(1973)}]{1973JQSRT..13..369A}
{Andersen}, T., \& {Soerensen}, G. 1973, \jqsrt, 13, 369

\bibitem[{{Aoki} {et~al.}(2007){Aoki}, {Honda}, {Sadakane}, \&
  {Arimoto}}]{2007PASJ...59L..15A}
{Aoki}, W., {Honda}, S., {Sadakane}, K., \& {Arimoto}, N. 2007, \pasj, 59, L15

\bibitem[{{Arnould}(1976)}]{1976A&A....46..117A}
{Arnould}, M. 1976, \aap, 46, 117

\bibitem[{{Ayres}(2010)}]{2010ApJS..187..149A}
{Ayres}, T.~R. 2010, \apjs, 187, 149

\bibitem[{{Biemont} {et~al.}(1983){Biemont}, {Grevesse}, {Hannaford}, {Lowe},
  \& {Whaling}}]{1983ApJ...275..889B}
{Biemont}, E., {Grevesse}, N., {Hannaford}, P., {Lowe}, R.~M., \& {Whaling}, W.
  1983, \apj, 275, 889

\bibitem[{Burbidge {et~al.}(1957)Burbidge, Burbidge, Fowler, \&
  Hoyle}]{RevModPhys.29.547}
Burbidge, E.~M., Burbidge, G.~R., Fowler, W.~A., \& Hoyle, F. 1957, Rev. Mod.
  Phys., 29, 547

\bibitem[{{Castelli} \& {Kurucz}(2003)}]{2003IAUS..210P.A20C}
{Castelli}, F., \& {Kurucz}, R.~L. 2003, in IAU Symposium, Vol. 210, Modelling
  of Stellar Atmospheres, ed. {N.~Piskunov, W.~W.~Weiss, \& D.~F.~Gray}, 20P--+

\bibitem[{{Clayton}(1968)}]{1968psen.book.....C}
{Clayton}, D.~D. 1968, {Principles of stellar evolution and nucleosynthesis}
  (New York: McGraw-Hill)

\bibitem[{{Cowan} {et~al.}(1991){Cowan}, {Thielemann}, \&
  {Truran}}]{1991PhR...208..267C}
{Cowan}, J.~J., {Thielemann}, F., \& {Truran}, J.~W. 1991, \physrep, 208, 267

\bibitem[{{Farouqi} {et~al.}(2009){Farouqi}, {Kratz}, \&
  {Pfeiffer}}]{2009PASA...26..194F}
{Farouqi}, K., {Kratz}, K., \& {Pfeiffer}, B. 2009, PASA, 26, 194

\bibitem[{{Farouqi} {et~al.}(2010){Farouqi}, {Kratz}, {Pfeiffer}, {Rauscher},
  {Thielemann}, \& {Truran}}]{2010ApJ...712.1359F}
{Farouqi}, K., {Kratz}, K., {Pfeiffer}, B., {Rauscher}, T., {Thielemann}, F.,
  \& {Truran}, J.~W. 2010, \apj, 712, 1359

\bibitem[{{Fran{\c c}ois} {et~al.}(2007){Fran{\c c}ois}, {Depagne}, {Hill},
  {Spite}, {Spite}, {Plez}, {Beers}, {Andersen}, {James}, {Barbuy}, {Cayrel},
  {Bonifacio}, {Molaro}, {Nordstr{\"o}m}, \& {Primas}}]{2007A&A...476..935F}
{Fran{\c c}ois}, P., {et~al.} 2007, \aap, 476, 935

\bibitem[{{Freiburghaus} {et~al.}(1999){Freiburghaus}, {Rembges}, {Rauscher},
  {Kolbe}, {Thielemann}, {Kratz}, {Pfeiffer}, \& {Cowan}}]{1999ApJ...516..381F}
{Freiburghaus}, C., {Rembges}, J.-F., {Rauscher}, T., {Kolbe}, E.,
  {Thielemann}, F.-K., {Kratz}, K.-L., {Pfeiffer}, B., \& {Cowan}, J.~J. 1999,
  \apj, 516, 381

\bibitem[{{Fr{\"o}hlich} {et~al.}(2006){Fr{\"o}hlich},
  {Mart{\'{\i}}nez-Pinedo}, {Liebend{\"o}rfer}, {Thielemann}, {Bravo}, {Hix},
  {Langanke}, \& {Zinner}}]{2006PhRvL..96n2502F}
{Fr{\"o}hlich}, C., {Mart{\'{\i}}nez-Pinedo}, G., {Liebend{\"o}rfer}, M.,
  {Thielemann}, F.-K., {Bravo}, E., {Hix}, W.~R., {Langanke}, K., \& {Zinner},
  N.~T. 2006, Physical Review Letters, 96, 142502

\bibitem[{{Gallagher} {et~al.}(2010){Gallagher}, {Ryan}, {Garc{\'{\i}}a
  P{\'e}rez}, \& {Aoki}}]{2010A&A...523A..24G}
{Gallagher}, A.~J., {Ryan}, S.~G., {Garc{\'{\i}}a P{\'e}rez}, A.~E., \& {Aoki},
  W. 2010, \aap, 523, A24+

\bibitem[{{Gregg} {et~al.}(2004){Gregg}, {Silva}, {Rayner}, {Valdes},
  {Worthey}, {Pickles}, {Rose}, {Vacca}, \& {Carney}}]{2004AAS...205.9406G}
{Gregg}, M.~D., {et~al.} 2004, in Bulletin of the American Astronomical
  Society, Vol.~36, American Astronomical Society Meeting Abstracts, 1496--+

\bibitem[{{Hoffman} {et~al.}(1996){Hoffman}, {Woosley}, {Fuller}, \&
  {Meyer}}]{1996ApJ...460..478H}
{Hoffman}, R.~D., {Woosley}, S.~E., {Fuller}, G.~M., \& {Meyer}, B.~S. 1996,
  \apj, 460, 478

\bibitem[{{Honda} {et~al.}(2007){Honda}, {Aoki}, {Ishimaru}, \&
  {Wanajo}}]{2007ApJ...666.1189H}
{Honda}, S., {Aoki}, W., {Ishimaru}, Y., \& {Wanajo}, S. 2007, \apj, 666, 1189

\bibitem[{{Honda} {et~al.}(2006){Honda}, {Aoki}, {Ishimaru}, {Wanajo}, \&
  {Ryan}}]{2006ApJ...643.1180H}
{Honda}, S., {Aoki}, W., {Ishimaru}, Y., {Wanajo}, S., \& {Ryan}, S.~G. 2006,
  \apj, 643, 1180

\bibitem[{{Ivans} {et~al.}(2006){Ivans}, {Simmerer}, {Sneden}, {Lawler},
  {Cowan}, {Gallino}, \& {Bisterzo}}]{2006ApJ...645..613I}
{Ivans}, I.~I., {Simmerer}, J., {Sneden}, C., {Lawler}, J.~E., {Cowan}, J.~J.,
  {Gallino}, R., \& {Bisterzo}, S. 2006, \apj, 645, 613

\bibitem[{{Johansson} {et~al.}(1994){Johansson}, {Joueizadeh}, {Litzen},
  {Larsson}, {Persson}, {Wahlstrom}, {Svanberg}, {Leckrone}, \&
  {Wahlgren}}]{1994ApJ...421..809J}
{Johansson}, S.~G., {et~al.} 1994, \apj, 421, 809

\bibitem[{{Kappeler} {et~al.}(1989){Kappeler}, {Beer}, \&
  {Wisshak}}]{1989RPPh...52..945K}
{Kappeler}, F., {Beer}, H., \& {Wisshak}, K. 1989, Reports on Progress in
  Physics, 52, 945

\bibitem[{{Kurucz}(1993{\natexlab{a}})}]{1993KurCD..13.....K}
{Kurucz}, R. 1993{\natexlab{a}}, ATLAS9 Stellar Atmosphere Programs and 2 km/s
  grid.~Kurucz CD-ROM No.~13.~ Cambridge, Mass.: Smithsonian Astrophysical
  Observatory, 1993., 13

\bibitem[{{Kurucz}(1993{\natexlab{b}})}]{1993KurCD..18.....K}
---. 1993{\natexlab{b}}, SYNTHE Spectrum Synthesis Programs and Line
  Data.~Kurucz CD-ROM No.~18.~Cambridge, Mass.: Smithsonian Astrophysical
  Observatory, 1993., 18

\bibitem[{{Kurucz}(2011)}]{2011CaJPh..89..417K}
{Kurucz}, R.~L. 2011, Canadian Journal of Physics, 89, 417

\bibitem[{{Lai} {et~al.}(2011){Lai}, {Smith}, {Bolte}, {Johnson}, {Lucatello},
  {Kraft}, \& {Sneden}}]{2011AJ....141...62L}
{Lai}, D.~K., {Smith}, G.~H., {Bolte}, M., {Johnson}, J.~A., {Lucatello}, S.,
  {Kraft}, R.~P., \& {Sneden}, C. 2011, \aj, 141, 62

\bibitem[{{Lawler} {et~al.}(2001){Lawler}, {Bonvallet}, \&
  {Sneden}}]{2001ApJ...556..452L}
{Lawler}, J.~E., {Bonvallet}, G., \& {Sneden}, C. 2001, \apj, 556, 452

\bibitem[{{Lodders}(2010)}]{2010ppc..conf..379L}
{Lodders}, K. 2010, in Principles and Perspectives in Cosmochemistry, ed.
  {A.~Goswami \& B.~E.~Reddy}, 379--+

\bibitem[{{Lugaro} {et~al.}(2003){Lugaro}, {Davis}, {Gallino}, {Pellin},
  {Straniero}, \& {K{\"a}ppeler}}]{2003ApJ...593..486L}
{Lugaro}, M., {Davis}, A.~M., {Gallino}, R., {Pellin}, M.~J., {Straniero}, O.,
  \& {K{\"a}ppeler}, F. 2003, \apj, 593, 486

\bibitem[{{Lundberg} {et~al.}(2010){Lundberg}, {Engstr{\"o}m}, {Hartman},
  {Nilsson}, {Palmeri}, {Quinet}, \& {Bi{\'e}mont}}]{2010JPhB...43h5004L}
{Lundberg}, H., {Engstr{\"o}m}, L., {Hartman}, H., {Nilsson}, H., {Palmeri},
  P., {Quinet}, P., \& {Bi{\'e}mont}, {\'E}. 2010, Journal of Physics B Atomic
  Molecular Physics, 43, 085004

\bibitem[{{Mayo} {et~al.}(2005){Mayo}, {Ortiz}, {Campos}, {Blagoev}, \&
  {Malcheva}}]{2005PhyS...72..142M}
{Mayo}, R., {Ortiz}, M., {Campos}, J., {Blagoev}, K., \& {Malcheva}, G. 2005,
  \physscr, 72, 142

\bibitem[{{Palmeri} {et~al.}(2009){Palmeri}, {Quinet}, {Fivet}, {Bi{\'e}mont},
  {Cowley}, {Engstr{\"o}m}, {Lundberg}, {Hartman}, \&
  {Nilsson}}]{2009JPhB...42p5005P}
{Palmeri}, P., {et~al.} 2009, Journal of Physics B Atomic Molecular Physics,
  42, 165005

\bibitem[{{Pellin} {et~al.}(2006){Pellin}, {Savina}, {Calaway}, {Tripa},
  {Barzyk}, {Davis}, {Gyngard}, {Amari}, {Zinner}, {Lewis}, \&
  {Clayton}}]{2006LPI....37.2041P}
{Pellin}, M.~J., {et~al.} 2006, in Lunar and Planetary Institute Science
  Conference Abstracts, Vol.~37, 37th Annual Lunar and Planetary Science
  Conference, ed. {S.~Mackwell \& E.~Stansbery}, 2041--+

\bibitem[{{Peterson}(2008)}]{2008STScINews...4.1P}
{Peterson}, R. 2008, Space Telescope Science Newsletter, 25, 24

\bibitem[{{Peterson} {et~al.}(2001){Peterson}, {Dorman}, \&
  {Rood}}]{2001ApJ...559..372P}
{Peterson}, R.~C., {Dorman}, B., \& {Rood}, R.~T. 2001, \apj, 559, 372

\bibitem[{{Roederer}(2011)}]{2011ApJ...732L..17R}
{Roederer}, I.~U. 2011, \apjl, 732, L17+

\bibitem[{{Roederer} {et~al.}(2010{\natexlab{a}}){Roederer}, {Cowan},
  {Karakas}, {Kratz}, {Lugaro}, {Simmerer}, {Farouqi}, \&
  {Sneden}}]{2010ApJ...724..975R}
{Roederer}, I.~U., {Cowan}, J.~J., {Karakas}, A.~I., {Kratz}, K., {Lugaro}, M.,
  {Simmerer}, J., {Farouqi}, K., \& {Sneden}, C. 2010{\natexlab{a}}, \apj, 724,
  975

\bibitem[{{Roederer} {et~al.}(2010{\natexlab{b}}){Roederer}, {Sneden},
  {Lawler}, \& {Cowan}}]{2010ApJ...714L.123R}
{Roederer}, I.~U., {Sneden}, C., {Lawler}, J.~E., \& {Cowan}, J.~J.
  2010{\natexlab{b}}, \apjl, 714, L123

\bibitem[{{Roederer} {et~al.}(2010{\natexlab{c}}){Roederer}, {Sneden},
  {Thompson}, {Preston}, \& {Shectman}}]{2010ApJ...711..573R}
{Roederer}, I.~U., {Sneden}, C., {Thompson}, I.~B., {Preston}, G.~W., \&
  {Shectman}, S.~A. 2010{\natexlab{c}}, \apj, 711, 573

\bibitem[{{Schwenke}(1998)}]{1998FaDi..109..321S}
{Schwenke}, D.~W. 1998, Faraday Discussions, 109, 321

\bibitem[{{Seeger} {et~al.}(1965){Seeger}, {Fowler}, \&
  {Clayton}}]{1965ApJS...11..121S}
{Seeger}, P.~A., {Fowler}, W.~A., \& {Clayton}, D.~D. 1965, \apjs, 11, 121

\bibitem[{{Sikstr{\"o}m} {et~al.}(2001){Sikstr{\"o}m}, {Pihlemark}, {Nilsson},
  {Litz{\'e}n}, {Johansson}, {Li}, \& {Lundberg}}]{2001JPhB...34..477S}
{Sikstr{\"o}m}, C.~M., {Pihlemark}, H., {Nilsson}, H., {Litz{\'e}n}, U.,
  {Johansson}, S., {Li}, Z.~S., \& {Lundberg}, H. 2001, Journal of Physics B
  Atomic Molecular Physics, 34, 477

\bibitem[{{Sneden} {et~al.}(2008){Sneden}, {Cowan}, \&
  {Gallino}}]{2008ARA&A..46..241S}
{Sneden}, C., {Cowan}, J.~J., \& {Gallino}, R. 2008, \araa, 46, 241

\bibitem[{{Sneden} {et~al.}(2003){Sneden}, {Cowan}, {Lawler}, {Ivans},
  {Burles}, {Beers}, {Primas}, {Hill}, {Truran}, {Fuller}, {Pfeiffer}, \&
  {Kratz}}]{2003ApJ...591..936S}
{Sneden}, C., {et~al.} 2003, \apj, 591, 936

\bibitem[{{Travaglio} {et~al.}(2004){Travaglio}, {Gallino}, {Arnone}, {Cowan},
  {Jordan}, \& {Sneden}}]{2004ApJ...601..864T}
{Travaglio}, C., {Gallino}, R., {Arnone}, E., {Cowan}, J., {Jordan}, F., \&
  {Sneden}, C. 2004, \apj, 601, 864

\bibitem[{{Whaling} \& {Brault}(1988)}]{1988PhyS...38..707W}
{Whaling}, W., \& {Brault}, J.~W. 1988, \physscr, 38, 707

\bibitem[{{Wickliffe} {et~al.}(1994){Wickliffe}, {Salih}, \&
  {Lawler}}]{1994JQSRT..51..545W}
{Wickliffe}, M.~E., {Salih}, S., \& {Lawler}, J.~E. 1994, \jqsrt, 51, 545

\bibitem[{{Woosley} \& {Howard}(1978)}]{1978ApJS...36..285W}
{Woosley}, S.~E., \& {Howard}, W.~M. 1978, \apjs, 36, 285

\bibitem[{{Xu} {et~al.}(2004){Xu}, {Persson}, {Svanberg}, {Blagoev},
  {Malcheva}, {Pentchev}, {Bi{\'e}mont}, {Campos}, {Ortiz}, \&
  {Mayo}}]{2004PhRvA..70d2508X}
{Xu}, H.~L., {et~al.} 2004, \pra, 70, 042508

\bibitem[{{Yong} {et~al.}(2008){Yong}, {Karakas}, {Lambert}, {Chieffi}, \&
  {Limongi}}]{2008ApJ...689.1031Y}
{Yong}, D., {Karakas}, A.~I., {Lambert}, D.~L., {Chieffi}, A., \& {Limongi}, M.
  2008, \apj, 689, 1031

\end{thebibliography}
